\documentclass[11pt,a4paper]{article}
\usepackage[latin9]{inputenc}
\usepackage{float}
\usepackage{amsmath}
\usepackage{graphicx}

\makeatletter

\pdfpageheight\paperheight
\pdfpagewidth\paperwidth

\providecommand{\tabularnewline}{\\}

\pdfoutput=1 

\usepackage{jinstpub}

\title{\boldmath Calibration and irradiation study of the BGO background monitor for the BEAST II experiment}

\author[a]{Y.T. Chen,}
\author[a,1]{J.C. Lin,\note{Corresponding author.}}
\author[a]{J.T. Chen,}
\author[a]{K.N. Chu,} 
\author[a]{K. Huang,}
\author[a]{S. Koirala,}
\author[a]{J.J. Liau,} 
\author[a]{F.H. Lin,}
\author[a]{K.P. Lin,}
\author[a]{J.G. Shiu,}
\author[a]{M.Z. Wang}

\affiliation[a]{Department of Physics, National Taiwan University,\\No. 1, Sec. 4, Roosevelt Rd., Taipei 10617, Taiwan (R.O.C.)}

\emailAdd{hd.5401@hotmail.com}

\abstract{Beam commissioning of the SuperKEKB collider began in 2016. The Beam
Exorcism for A STable experiment II (BEAST II) project is particularly
designed to measure the beam backgrounds around the interaction point
of the SuperKEKB collider for the Belle II experiment. We develop
a system using bismuth germanium oxide (BGO) crystals with optical
fibers connecting to a multianode photomultiplier tube (MAPMT) and
a field-programmable gate array (FPGA) embedded readout board for
monitoring the real-time beam backgrounds in BEAST II. The overall
radiation sensitivity of this system is estimated to be $(2.20\pm0.26)\times10^{-12}$
Gy/ADU (analog-to-digital unit) with the standard 10 m fibers for
transmission and the MAPMT operating at 700 V. Our $\gamma$-ray irradiation
study of the BGO system shows that the exposure of BGO crystals to
$^{60}$Co $\gamma$-ray doses of 1 krad has led to immediate
light output reductions of 25\textendash 40\%, and the light outputs
further drop by 30\textendash 45\% after the crystals receive doses
of 2\textendash 4 krad. Our findings agree with those of the previous studies on the radiation
hard (RH) BGO crystals grown by the low thermal gradient Czochralski
(LTG Cz) technology. The absolute dose from the BGO system is also
consistent with the simulation, and is estimated to be about 1.18
times the equivalent dose. These results prove that the BGO system is 
able to monitor the background dose rate in real time under extreme high 
radiation conditions. This study concludes that the BGO system is reliable for the beam background study in BEAST II.}


\arxivnumber{1705.00312} 

\makeatother

\begin{document}
\maketitle \flushbottom

\section{Introduction}

The SuperKEKB collider \cite{RN422} at KEK, Japan is designed to
supply high-luminosity $e^{+}/e^{-}$ collisions for the Belle II
experiment \cite{RN392,RN421} (the successor to the Belle experiment
\cite{RN425}). Its first beam circulation in 2016 brings forth a
new era of accelerator operation. The Beam Exorcism for A Stable experiment
II (BEAST II) project is dedicated to study beam-induced backgrounds
prior to the Belle II full installation and commissioning. This is
essential for beam commissioning and the protection of the real Belle
II detector from unexpected high radiation. The HEP group of National
Taiwan University (NTUHEP) has developed a compact system for monitoring
the real-time beam backgrounds in BEAST II. The measured beam backgrounds
are centralized in the BEAST II DAQ (data acquisition) system and
fed back to the SuperKEKB control center. 

The BGO system adopts bismuth germanate Bi$_{4}$Ge$_{3}$O$_{12}$
(BGO) crystal sensors for the detection of charged particles and photon
background. BGO is a well-known material extensively used as scintillators.
Since the scintillation property of BGO was discovered \cite{RN439},
its fine $e/\gamma$ energy resolution, high density, large refractive
index, and non-hygroscopic nature have promoted its widely application
in high energy physics, nuclear physics, positron emission tomography
(PET), and many other fields. However, the high sensitivity to radiation
damage of BGO crystal upon irradiation has limited its use in high
energy applications. Therefore, several works have been conducted
to study the radiation damage to BGO crystals. Early studies by Kobayashi
et al. (1983) \cite{RN438} reported that the extent to which a crystal
is damaged is related to its impurity concentration. Wei et al. (1990)
discovered that the radiation resistance of BGO crystals is improved
by europium doping \cite{RN437}. Yanovsky et al. (1991), on the other
hand, revealed that BGO crystals from certain manufacturers show very
high radiation hardness even up to 100 Mrad of irradiation and without
any photochromic effect \cite{RN436}. Zhu et al. (1995) also showed
that doped BGO crystals could suffer damage by radiation even at 1
krad \cite{RN435}. Georgii et al. (1998) further performed systematic
radiation damage studies on BGO crystals grown by different manufacturers
\cite{RN432}.

The low thermal gradient Czochralski (LTG Cz) technique, developed
by Nikolaev Institute of Inorganic Chemistry (NIIC) of the Siberian
Branch of the Russian Academy of Sciences, is commonly used for growing
BGO crystals \cite{RN434}. Compared with the conventional Czochralski
(Cz) technique and the Bridgman method, the LTG Cz technique reduces
the temperature gradients, and the solidification front becomes fully
faceted. It improves the quality of the crystals. In addition, under
the LTG Cz condition, the sensitive properties of crystals to the
growth mechanism allow one to control the radiation hardness of BGO
crystals by choosing appropriate growth conditions. As a result, radiation
hard (RH) crystals are grown. The LTG Cz technology was developed
in the NIIC as a routine way to grow nearly perfect undoped RH BGO
crystals.

The first large scale application of the RH BGO crystals grown by
the LTG Cz technology was the extreme forward calorimeter (EFC) \cite{RN429}
of the Belle detector at KEK B-factory. Our group has constructed
this calorimeter to improve coverage of small angles around the beam
pipe of the Belle detector, and has surveyed the effects of radiation
damage and recovery of undoped RH BGO crystals \cite{RN428,RN433,RN431,RN430}.
The results revealed that the BGO crystals produced using the same
LTG Cz growth technique show two distinct types of behavior under
irradiation and relaxation. These crystals from mass production \cite{RN428}
(the EFC crystals) are less radiation-hard than the crystals produced
in small quantity (the sample crystals) \cite{RN433,RN431,RN430}.
The former also show considerably slower self-recovery compared to
the latter. Similar studies have also been carried out by NIIC \cite{RN426,RN427,RN424},
and the results also revealed that the BGO crystals grown by the same
LTG Cz technology exhibit two distinct types of characteristics. The
crystals termed as L-type show weak monotonous degradation of light
output upon irradiation and fast relaxation. They remain colorless
even after receiving doses of several Mrad. The crystals termed as
N-type show strong abrupt degradation of light output upon irradiation
and slow relaxation. They become yellow tinted after receiving 1 krad
$\gamma$-radiation doses. These findings correspond with the results
reported by our group. However, the reasons for the distinct features
of the BGO crystals grown by the same LTG Cz technique are not clear
yet. Although a considerable difference in electrical self-conductivity
is discovered for L- and N-type crystals, there is no noticeable difference
in the impurity content between the two types of crystals.

We reuse the scintillating BGO crystals from EFC, and apply them as
the sensors of the BGO detector system. We define the radiation sensitivity
for one channel of the detector as the calibrated value for the ionizing
radiation dose of the BGO crystal normalized to one ADU (analog-to-digital
unit) count. In order to determine the radiation sensitivities of
the BGO system, the system was first calibrated by a LED source to
determine the gain factors of the multianode photomultiplier tube
(MAPMT) in units of ADU per photoelectron (p.e.). The p.e yield (p.e./GeV;
defined as $Y_{{\rm p.e.}}$) of this system is then determined based
on the comparison between the minimum ionizing cosmic ray data and
the results of a Geant4 simulation. A special test with the proposed
configuration for BEAST II was conducted at the Institute of Nuclear
Energy Research (INER), Longtan, Taiwan by a strong $^{60}$Co irradiation
facility to examine the performance of the system. The results obtained
from the BGO system are compared with those obtained from dosimeters
and from a Geant4 simulation, respectively. 

In this work, we report the design of our system, the calibration
procedures, and the results of the irradiation study. The results
of the irradiation study are expected to agree with the those of the
previous studies performed on the RH BGO crystals grown by the LTG
Cz technology. This can demonstrate the success of the BGO detector
system and ensure that the data provided by our system for BEAST II
are reliable.

\section{BGO detector system}

\begin{figure}[H]
\centering\includegraphics[width=9cm]{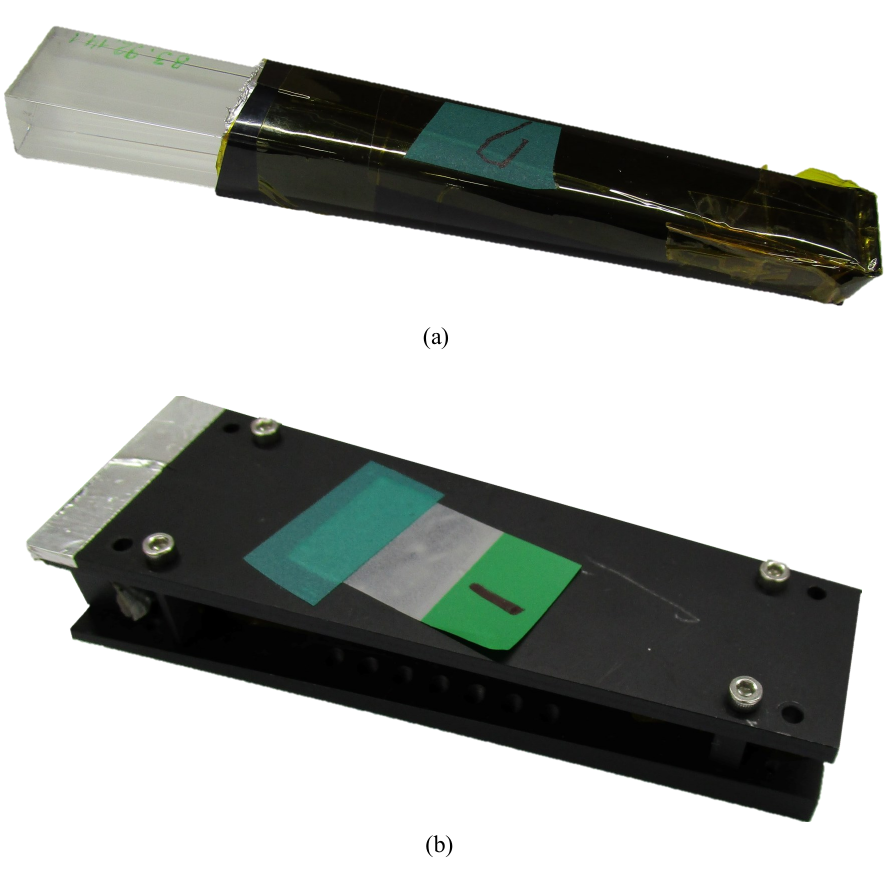}

\caption[Photograph of one of the BGO crystal sensors and the supporting structure.]{Photograph of one of the BGO crystal sensors and the supporting structure.
(a) A BGO crystal half-pulled out from the wrapping material for demonstration
purpose. (b) Supporting structure of the BGO crystal sensors.}

\label{fig:Photograph of one of the BGO crystal sensors and the supporting structure}
\end{figure}

\begin{figure}[H]
\centering\includegraphics[width=9cm]{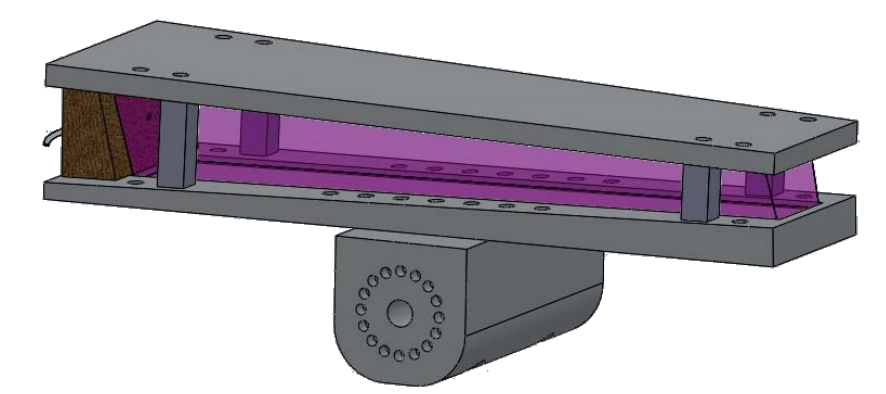}

\caption[CAD rendering of one of the BGO crystal sensors and the supporting
structure.]{CAD rendering of one of the BGO crystal sensors and the supporting
structure.}

\label{fig:CAD rendering of one of the BGO crystal sensors and the supporting structure}
\end{figure}

We choose BGO as the electromagnetic shower medium for beam-induced
radiation, as shown in figure \ref{fig:Photograph of one of the BGO crystal sensors and the supporting structure}
and figure \ref{fig:CAD rendering of one of the BGO crystal sensors and the supporting structure}.
The dimension of each BGO crystal is about $2\times2\times13$ cm$^{3}$
and has a mass of approximately 0.3135 kg. Each BGO crystal is wrapped
with a 100 $\mu$m thick Teflon sheet and a 25 $\mu$m thick aluminized
Mylar sheet to obtain the maximum light-collection efficiency. Some
supporting structures made of nonmagnetic materials are designed to
protect the crystals.

\begin{figure*}[tp]
\centering\includegraphics[width=14cm]{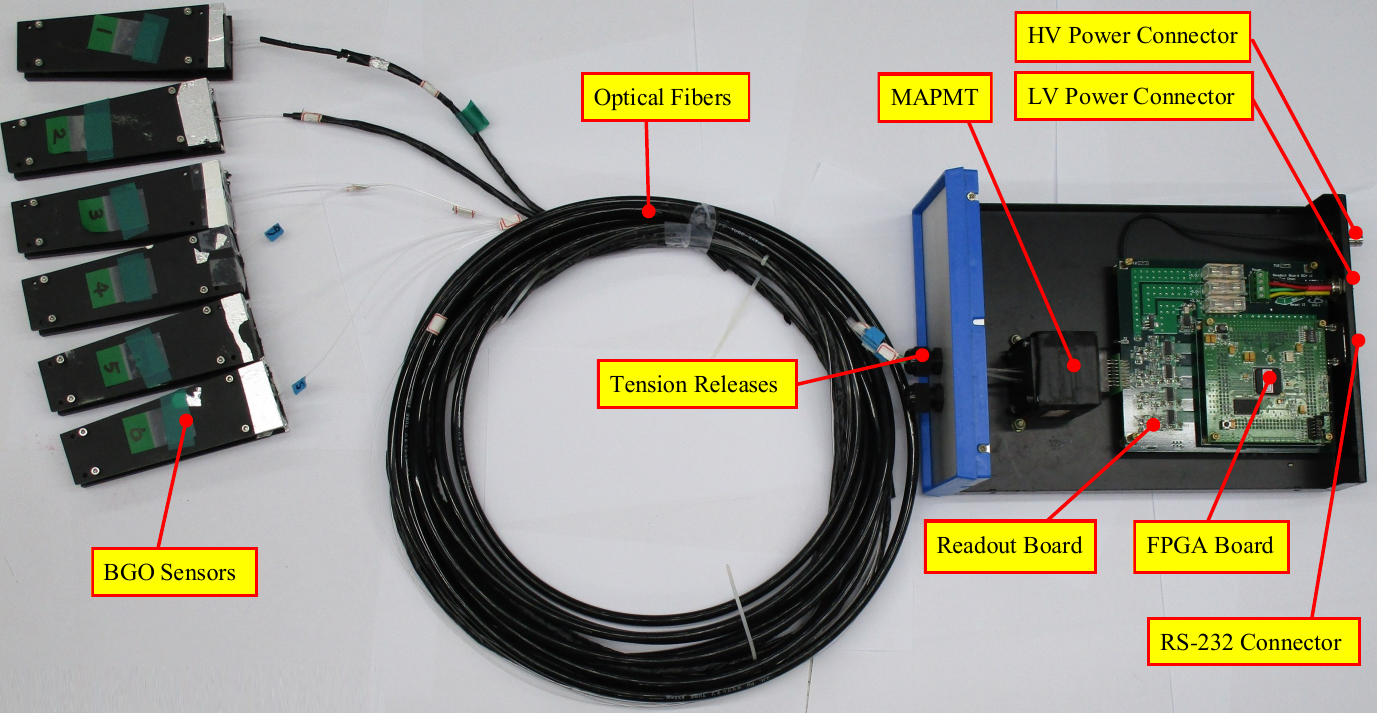}

\caption[Layout of the BGO detector system.]{Layout of the BGO detector system.}
\label{fig:Layout of the BGO detector system}
\end{figure*}

\begin{figure}[tbph]
\centering\includegraphics[width=9cm]{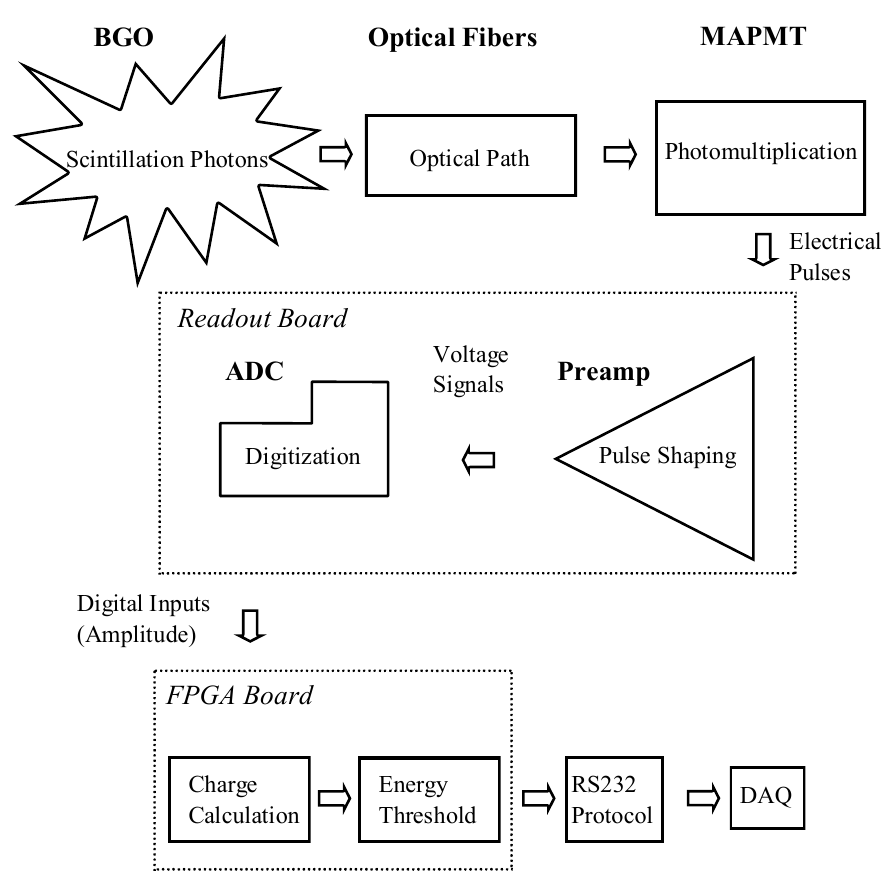}\caption[Schematic of the BGO detector system.]{Schematic of the BGO detector system.}
\label{fig:Schematic of the BGO detector system}
\end{figure}

The layout and the schematic of the detector are shown in figure \ref{fig:Layout of the BGO detector system}
and figure \ref{fig:Schematic of the BGO detector system}, respectively.
The BGO system can at most operate with eight BGO sensors. To simplify
the design and avoid radiation damage to the electronics, there are
no active components on the side of the BGO crystals. The signal from
each BGO crystal is conducted to a Hamamatsu H7546 MAPMT via one 10
m Eska SK-40 plastic fiber (1 mm in diameter). A MAPMT mask points
each fiber to the center of the corresponding MAPMT pixel. Some light-tight
treatments are applied to the fibers and their connections to prevent
leakage of light from the environment. The MAPMT, which has a spectral
range of 300\textendash 650 nm, converts the scintillation light to
charge signals. The charge signals from the MAPMT are connected to
the readout electronics system. The readout system and the MAPMT are
encased in a electrically grounded DAQ box for portability and electromagnetic
noise shielding. 

The readout system consists of two parts: an eight-channel readout
board and a Xilinx 3A/3AN field-programmable gate array (FPGA) board.
The FPGA is plugged on top of the readout board, and the firmware
developed at Xilinx ISE (Integrated Synthesis Environment) is written
by VHDL (VHSIC hardware description language). The readout board inherits
the design from the NuTel experiment \cite{RN423} with some modifications
to improve its noise immunity and performance of shaping signals.
Each channel of the readout board consists of a preamplifier circuit
and a 10-bit pipeline ADC (analog-to-digital converter) chip. The
preamplifier reconverts the analog charge signal to analog voltage
signal by shaping the pulses through the RC circuit with a 187 ns
shaping time, and then the following ADC digitizes the analog voltage
signal. With the specific 187 ns shaping time, the input pulse height
will drop by about 1/8 after one clock (25 ns) if there are no new
input changes. The excess of that will be regarded as the input charges
during that clock. Hence. the input charges are calculated by using
the following algorithm: \begin{subequations}

\begin{align}
Q_{in}(t) & =A(t)-A(t-1)\times e^{\left(-\frac{25}{187}\right)},\label{eq:y:1-1}\\
 & \approx A(t)-A(t-1)\times\frac{7}{8},\label{eq:y:2-1}
\end{align}
\end{subequations}where $A(t)$ is the digital signal in the FPGA
(in ADUs), $Q_{in}(t)$ is the accumulated input charges in the increment
of the 25 ns clock period (in ADUs). After converting the pulse heights
to the input charges, the FPGA delivers the acquired data for calculating
the accumulated radiation doses to the PCs via a RS-232 serial cable
at a data rate of 65.91 Hz without dead time. The computer program
processes the raw data and checks the validity. The processed data
include the timestamps, the accumulated radiation doses, the severity
of ADC overflow, and the checksums. The computer program displays
the background conditions on the screen in real time, and stores the
processed data on the disk.

\section{Calibration methods}

\subsection{Calibration of gains}

\begin{figure}[H]
\centering \includegraphics[width=10cm]{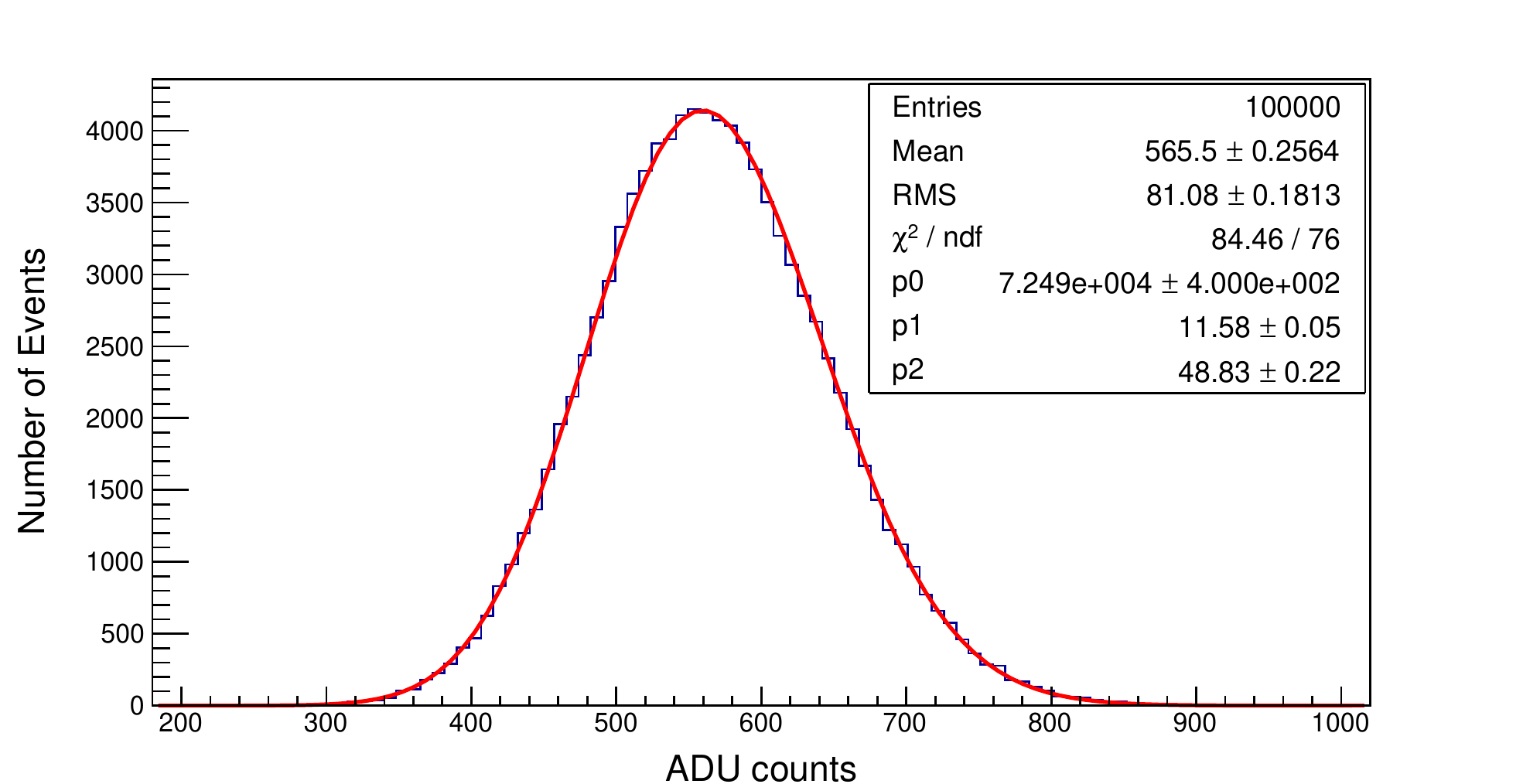}

\caption[A typical charge spectrum of the MAPMT operating at 700 V.]{A typical charge spectrum of the MAPMT operating at 700 V.}
\label{fig:A typical charge spectrum of the MAPMT operating at 700 V} 
\end{figure}

A LED calibration procedure was used to evaluate the gain factors
of the MAPMT. We shone light pulses from a 450 nm blue-light LED onto
each MAPMT pixel. The intensity of the blue light pulses can be adjusted
by setting the input voltage of the LED. The number of p.e. induced
by each incident light pulse in the cathode follows a Poisson distribution
\cite{RN440}:

\begin{equation}
P(C)=\frac{N^{\frac{C}{G}}}{(\frac{C}{G})!}e^{-N},\label{eq:1}
\end{equation}
where $P$ is the probability of having $N$ photoelectrons, $C$
is the number of charges (in units of ADUs), $N$ is the number of
p.e., and $G$ is the gain factor (in units of ADU/p.e.). The obtained
charge spectra are fitted with eq. \ref{eq:1}, as shown in figure
\ref{fig:A typical charge spectrum of the MAPMT operating at 700 V}.

\subsection{Calibration of photoelectron yield}

\begin{figure}[H]
\centering\includegraphics[width=9cm]{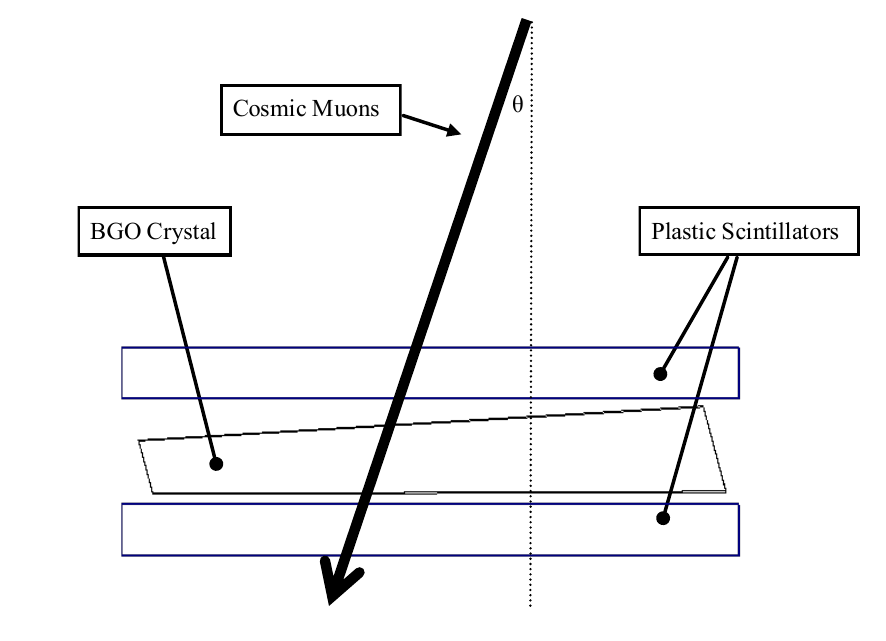}

\caption[Schematic of the trigger system for cosmic muon MIPs.]{Schematic of the trigger system for cosmic muon MIPs.}
\label{fig:Schematic of the trigger system for cosmic muon MIPs}
\end{figure}

$Y_{{\rm p.e.}}$ of this system is obtained based on the comparison
between the minimum ionizing particles (MIPs) data and the results
of the simulation. The schematic of the trigger system for cosmic
muon MIPs is shown in figure \ref{fig:Schematic of the trigger system for cosmic muon MIPs}.
A PMT was coupled to a BGO crystal through fifty 10-cm Eska SK-40
fibers, and the BGO crystal was sandwiched between two horizontal
plastic scintillators. The trigger algorithm was set to ensure that
the deposited energy of the BGO crystal and that of each plastic scintillator
are larger than the threshold energy\textemdash the cosmic-ray muon
should hit the BGO crystal and the two scintillators. In the simulation,
the energy of each cosmic-ray muon was set at 10 GeV, and the rate
of which was proportional to $\cos^{2}(\theta)$, where $\theta$
is the incident angle deviated from the cosmic rays. The spectra obtained
are fitted with Landau convoluted with Gaussian. $Y_{{\rm p.e.}}$
is then determined by comparing the position of the two MIP peaks:

\begin{equation}
Y_{{\rm p.e.}}=\frac{MP_{data}}{MP_{simulation}}\times\frac{Q_{ratio}}{N_{fiber}},
\end{equation}
where
\begin{itemize}
\item $MP_{data}$ is the most probable value of the energy deposition spectrum
of cosmic muon MIPs obtained from the BGO system.
\item $MP_{simulation}$ is the most probable value of the energy deposition
spectrum of cosmic muon MIPs obtained from the simulation.
\item $Q_{ratio}$ is the quantum efficiency of the MAPMT with respect to
that of the PMT of this calibration.
\item $N_{fiber}$ is the number of fibers of the trigger system.
\end{itemize}
To determine $Q_{ratio}$, we designed a special test. We shone light
pulses with the same amount of energy onto the MAPMT and the PMT.
$Q_{ratio}$ is determined by the following formula:

\begin{eqnarray}
Q_{ratio} & = & \frac{N_{MAPMT}}{N_{PMT}}\times\frac{A_{PMT}}{A_{MAPMT}},
\end{eqnarray}
where $N_{MAPMT}$ is the number of p.e. induced by each incident
light pulse onto the MAPMT pixel; $N_{PMT}$ is the number of p.e.
induced by each incident light pulse onto the PMT pixel; $A_{PMT}$
is the photon-receiving area of the PMT; $A_{MAPMT}$ is the photon-receiving
area of the MAPMT. To examine the alignment of the MAPMT mask, we
measured $Q_{ratio}$ with and without the MAPMT mask, respectively.

\section{Irradiation study}

\begin{figure*}[tp]
\centering\includegraphics[width=15cm]{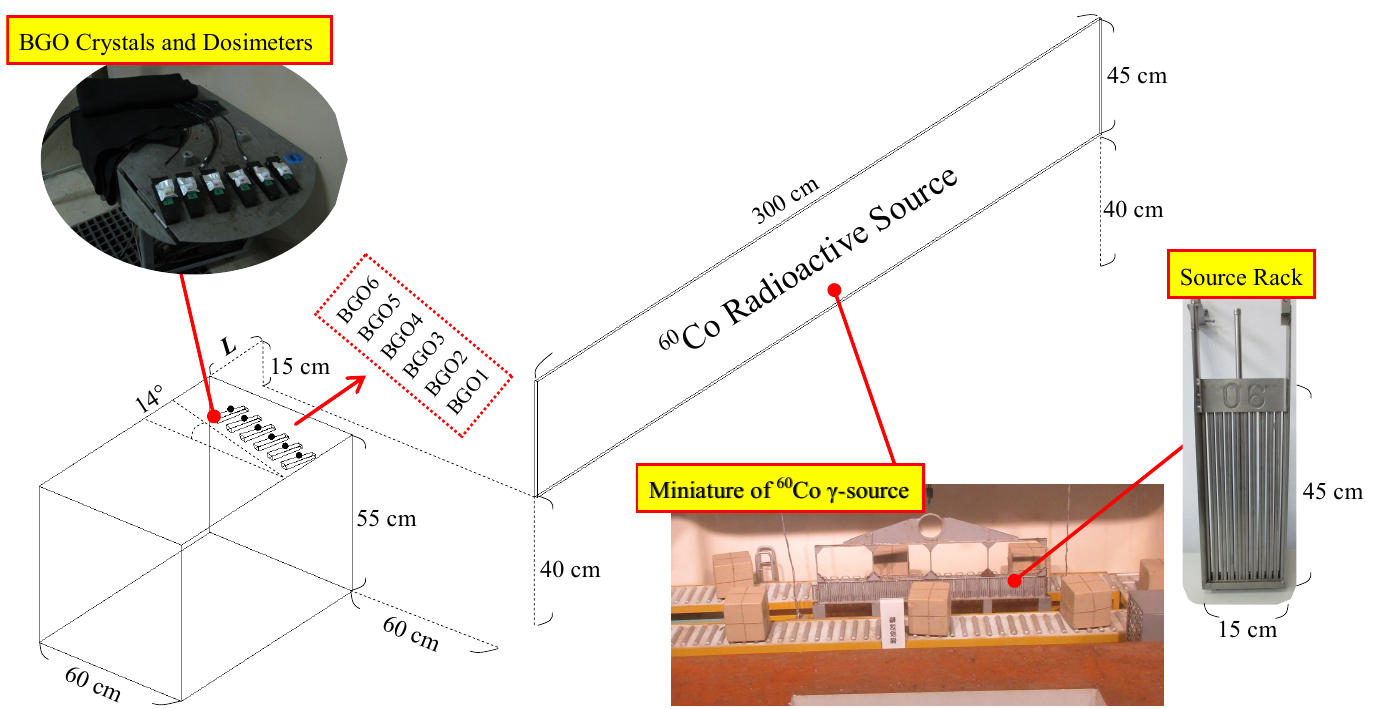}\caption[Schematic of the $\gamma$-ray irradiation at INER.]{Schematic of the $\gamma$-ray irradiation at INER. The entire $\gamma$-source
consisted of 20 source racks, and each source rack contained 20 sticks
with sealed $^{60}$Co. The diameter of each stick was 10 mm. The
distance between two adjacent BGO crystals was about 7.2 cm. $L$
in this schematic is defined as the distance between the source and
the BGO crystals.}
\label{fig:Schematic of the gamma-ray irradiation study at INER} 
\end{figure*}

Only six of the eight BGO crystals were mounted on the BGO system
for the irradiation test because the other two had been sent to the
University of Hawaii for further tests. Some commercial alanine dosimeters
(AWM230 from A. Wieser Messtechnik, batch 00303/M5) prepared by INER
were placed on top of each BGO crystal. The irradiation source was
$^{60}$Co with a dimension of about $300\times45\times1$ cm$^{3}$.
The activity of the source was around $1.11\times10^{14}$ Bq. Schematic
of the experimental setup for the $\gamma$-ray irradiation is shown
in figure \ref{fig:Schematic of the gamma-ray irradiation study at INER}.

We conducted five irradiation steps with different irradiation conditions.
In this study, the absolute dose is defined as the dose absorbed by
the sample, and the equivalent dose is defined as the amount of dose
that would have been absorbed by water if placed at the same position.
The BGO system monitored the light output (p.e. rate) of each crystal
in real time to determine to the absolute dose; the dosimeters measured
the equivalent dose for each crystal. Besides, the Geant4 simulation
provided simulated absolute doses for the whole irradiation procedure. 

In each irradiation step, the source would quickly rise from water
until it reached the highest point, and then remained stable for 15
minutes. In the first step, a bare BGO crystal was put near the source
to examine the light-tightness of the system. The scintillation light
from the bare crystal functioned as a light pollution source from
the environment. The bare crystal glowed brightly during the irradiation,
and appeared to be yellow tilted and warmer after removal. In the
last step, the table was rotated 90$^{\circ}$ counterclockwise to
have the crystals irradiated transversely. The details of each step
are given in table \ref{table:Log of irradiation steps}. We first
studied the dependence of light outputs on the elapsed time and the
absorbed dose, respectively. Then the absolute doses in each step
measured by the BGO system were compared with those obtained from
dosimeters and from the simulation, respectively. The absolute doses
from the BGO system are obtained based on the following assumptions: 
\begin{enumerate}
\item The dose rate in each step is a step function with a duration of exactly
15 minutes corresponding to the stable periods of the source.
\item The degradation of light output in each step is accumulated to the
next step without recovery.
\end{enumerate}
The whole irradiation procedure lasted for $\sim$2.5 h, and the total
accumulated dose of each crystal was about 4.5 krad. The dosimeters
were 10 mm in length and 4.8 mm in diameter, and their suitable range
of dose measurement was 1\textendash 100 Gy, which meets our dose
measurement requirements. The irradiation test was performed in a
darkroom, and the system was covered with block cloth to strengthen
the light-tightness. A pedestal run was carried out to check the noise
level from electronics and environmental radiation prior to the test.

Although the experimental setup was covered with block cloth, the
connections between the fibers and the crystals were not glued with
silicon, allowing in-situ adjustments. Inevitably, the scintillation
light of the bare crystal in the first step might contribute some
extra light outputs to the system if there was light leakage. The
light outputs due to the bare crystal were removed carefully when
we calculated the absolute doses.

\begin{table*}
\centering\caption[Log of irradiation steps]{Log of irradiation steps.}
\smallskip{}

\label{table:Log of irradiation steps} %
\begin{tabular}{|c|c|c|c|}
\hline 
Step & Start time (minutes)  & $L$ (m)  & Note\tabularnewline
\hline 
1  & 8 & 3.0 & Light-tightness test\tabularnewline
2  & 36  & 3.0 & \tabularnewline
3  & 74 & 2.0 & \tabularnewline
4  & 102  & 1.5 & \tabularnewline
5  & 132 & 2.0 & Irradiated transversely\tabularnewline
\hline 
\end{tabular}
\end{table*}

\section{Results and discussion}

\subsection{Calibration}

\subsubsection{Gain factors of the MAPMT}

The gain factors of the MAPMT operating at 700 V are summarized in
table \ref{table:Gain factors of the MAPMT operating at 700 V}. The
results agree with the values in the data sheet.

\begin{table}[h]
\centering\caption[Gain factors of the MAPMT operating at 700 V.]{Gain factors of the MAPMT operating at 700 V.}
\smallskip{}
\label{table:Gain factors of the MAPMT operating at 700 V} %
\begin{tabular}{|c|c|}
\hline 
BGO channel & Gain (ADU/p.e.)\tabularnewline
\hline 
CH1  & $10.90\pm0.40$\tabularnewline
CH2  & $13.18\pm0.43$\tabularnewline
CH3  & $10.82\pm0.42$\tabularnewline
CH4  & $12.65\pm0.54$\tabularnewline
CH5  & $10.58\pm0.46$\tabularnewline
CH6  & $11.93\pm0.64$\tabularnewline
CH7  & $10.51\pm0.42$\tabularnewline
CH8  & $11.62\pm0.61$\tabularnewline
\hline 
\end{tabular}
\end{table}

\subsubsection{Photoelectron yield}

We assume that the alignment of the MAPMT mask is perfect enough to
centralize and restrict the light to the corresponding pixels without
significant loss. The results of the measurements of $Q_{ratio}$
with and without the MAPMT mask are listed below:
\begin{description}
\item [{With the MAPMT mask:}] With optical fibers for transmission. One
end of the fiber was connected to the PMT pixel with a special fixture.
$A_{MAPMT}$ and $A_{PMT}$ both were 1 mm$^{2}$, the cross section
area of the fibers. $Q_{ratio}$ is determined to be $1.438\pm0.014$.
\item [{Without the MAPMT mask:}] Without optical fibers for transmission.
$A_{PMT}$ was reduced to 3.799 mm$^{2}$ by a collimator; $A_{MAPMT}$
was 4 mm$^{2}$, the pixel area of the MAPMT. $Q_{ratio}$ is determined
to be $1.47\pm0.11$.
\end{description}
There is no significant difference between the values of $Q_{ratio}$
measured under the two different conditions, justifying the previous
assumption. The alignment of the MAPMT mask is very good. $Q_{ratio}$
is then determined to be $1.45\pm0.08$.

\begin{figure}[H]
\centering \includegraphics[width=9.5cm]{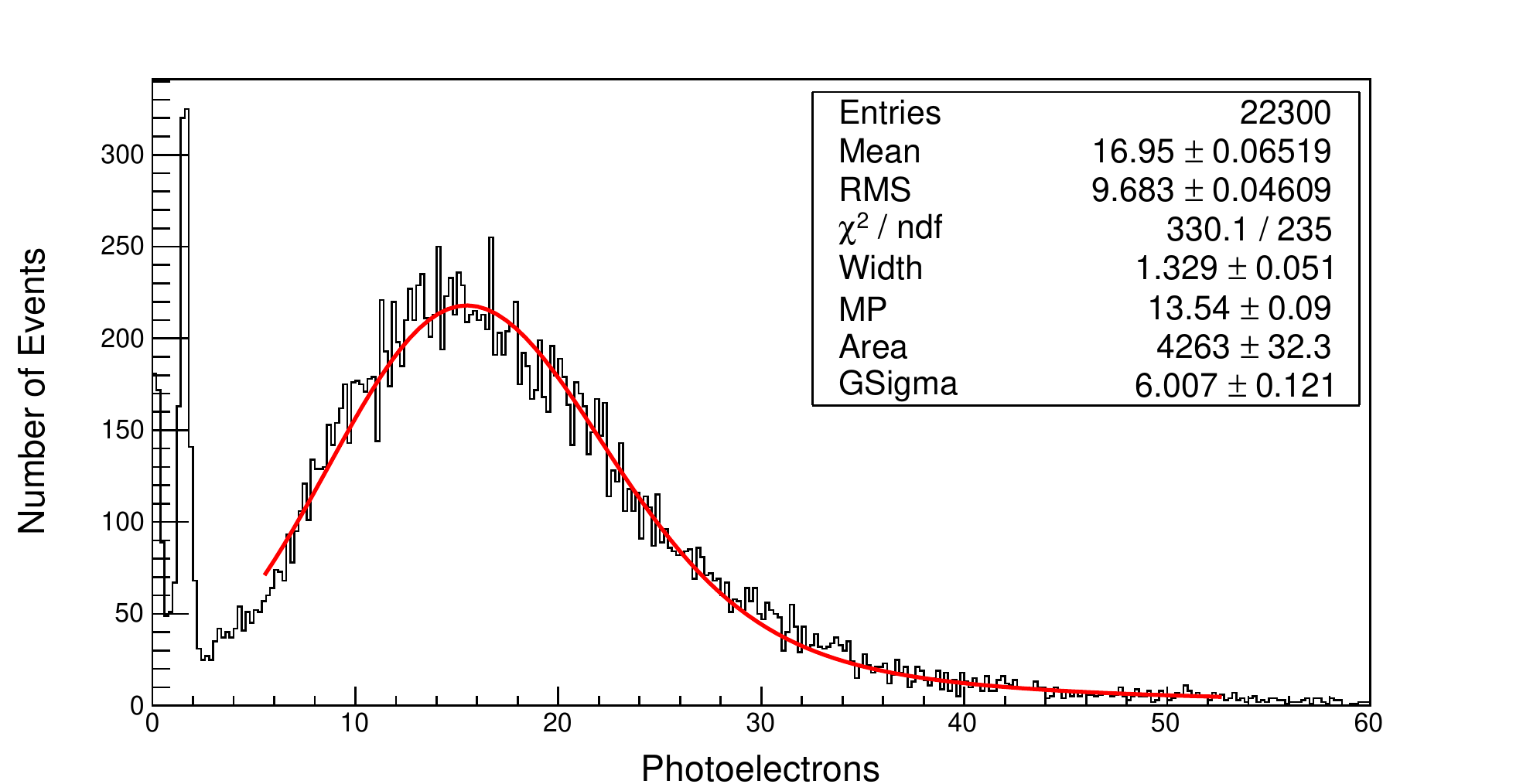}\caption[Energy deposition spectrum of cosmic muon MIPs obtained from BGO system.]{Energy deposition spectrum of cosmic muon MIPs obtained from BGO
system. }
\label{fig:Energy deposition spectrum of cosmic muon MIPs obtained from BGO system}
\end{figure}

\begin{figure}[H]
\centering\includegraphics[width=9.5cm]{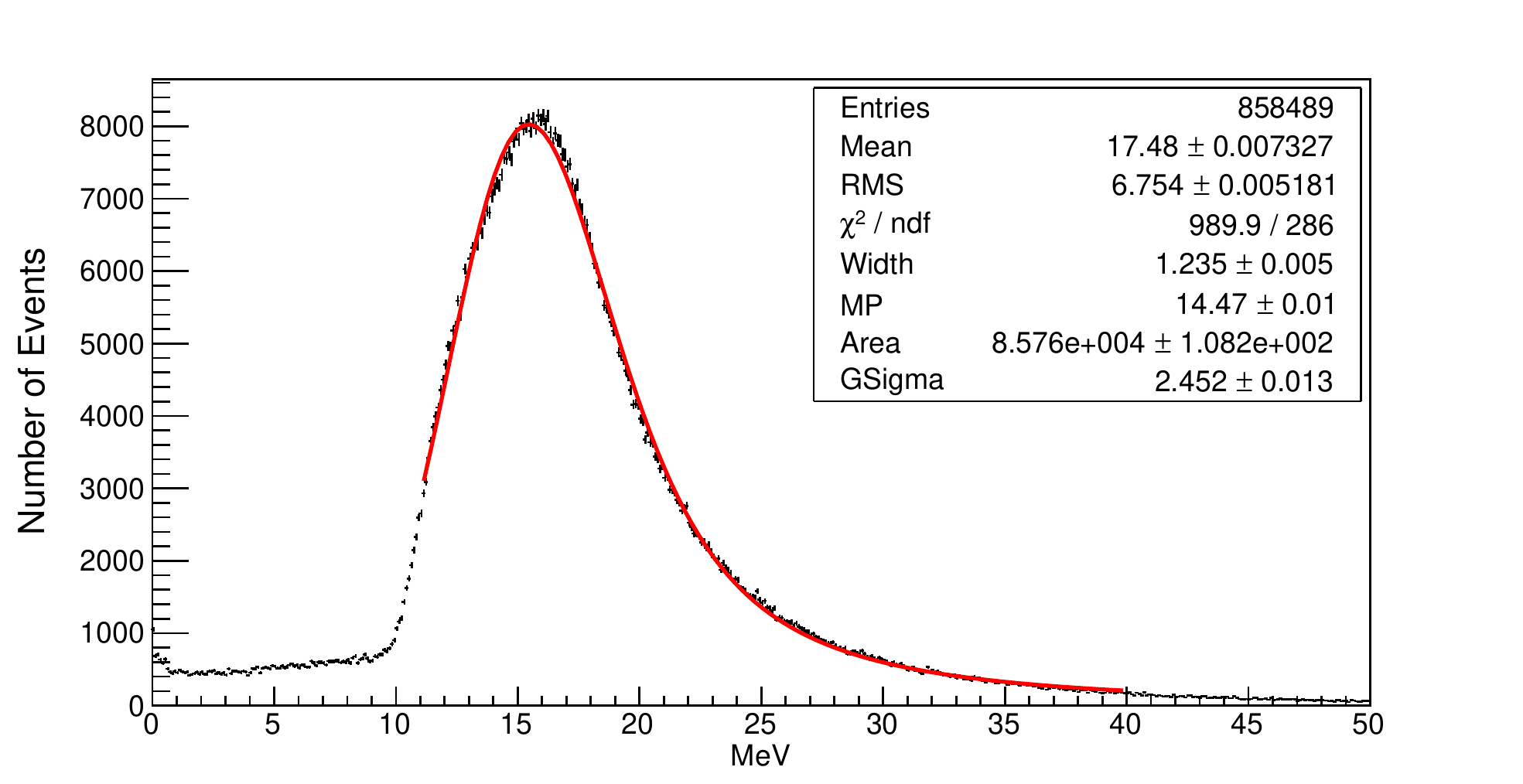}\caption[Energy deposition spectrum of cosmic muon MIPs obtained from simulation.]{Energy deposition spectrum of cosmic muon MIPs obtained from simulation. }
\label{fig:Energy deposition spectrum of cosmic muon MIPs obtained from simulation}
\end{figure}

Figure \ref{fig:Energy deposition spectrum of cosmic muon MIPs obtained from BGO system}
and figure \ref{fig:Energy deposition spectrum of cosmic muon MIPs obtained from simulation}
show the energy deposition spectra of cosmic muon MIPs obtained from
the BGO system and the simulation, respectively, where $MP_{data}$
is $13.54\pm8.66$ p.e. and $MP_{simulation}$ is $14.47\pm0.72$
MeV. $MP_{simulation}$ is consistent with the $dE/dx$ $(\sim$9
MeV/cm) of MIPs passing through BGO. Therefore, $Y_{{\rm p.e.}}$
is determined to be $27.17\pm1.51$ p.e./GeV with ideally zero meter
fibers for transmission. However, with the standard 10 m fibers for
transmission, the signals would attenuate to $(74.58\pm3.85)\%$,
according to our measurements. As a results, $Y_{{\rm p.e.}}$ is
determined to be $20.27\pm1.54$ p.e./GeV. with the standard 10 m
fibers for transmission.

\subsubsection{Radiation sensitivity}

The radiation sensitivities of the BGO detector system are given in
table \ref{table:Radiation sensitivities of BGO detector system with the MAPMT operating at 700 V}.
The overall radiation sensitivity of this system is estimated to be
$(2.20\pm0.26)\times10^{-12}$ Gy/ADU with the standard 10 m fibers
for transmission and the MAPMT operating at 700 V.

\noindent 
\begin{table}[h]
\centering\caption[Radiation sensitivities of BGO detector system with the MAPMT operating
at 700 V.]{Radiation sensitivities of BGO detector system with the MAPMT operating
at 700 V.}
\smallskip{}
\label{table:Radiation sensitivities of BGO detector system with the MAPMT operating at 700 V}
\begin{tabular}{|c|c|}
\hline 
BGO channel  & Sensitivity (Gy/ADU) \tabularnewline
\hline 
CH1  & $(2.31\pm0.20)\times10^{-12}$\tabularnewline
CH2  & $(1.91\pm0.16)\times10^{-12}$\tabularnewline
CH3  & $(2.33\pm0.20)\times10^{-12}$\tabularnewline
CH4  & $(1.99\pm0.17)\times10^{-12}$\tabularnewline
CH5  & $(2.38\pm0.21)\times10^{-12}$\tabularnewline
CH6  & $(2.11\pm0.20)\times10^{-12}$\tabularnewline
CH7  & $(2.40\pm0.21)\times10^{-12}$\tabularnewline
CH8  & $(2.17\pm0.20)\times10^{-12}$\tabularnewline
\hline 
\end{tabular}
\end{table}

\subsection{Irradiation study}

The pedestal run before the irradiation test showed that the noise
from the electronics and the environmental radiation can be negligible.
Because of some optical-path problems, the response of BGO-4 is only
about one fourth of the other signals. Hence, BGO-4 is excluded from
our analyses.

\begin{figure}[h]
\centering\includegraphics[width=9cm]{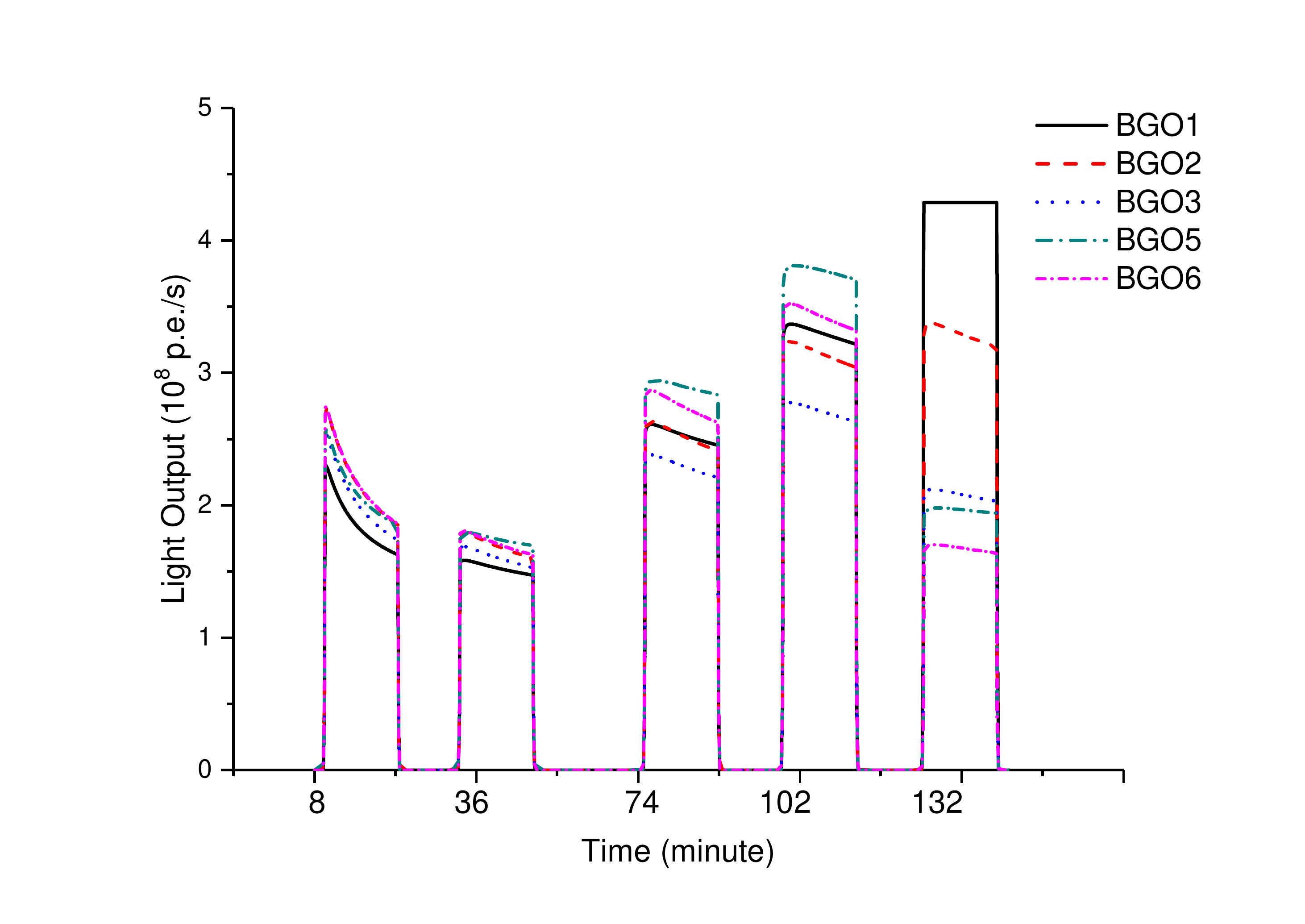} 

\caption[Dependence of the light output on the elapsed time in the irradiation
test.]{Dependence of the light output on the elapsed time in the irradiation
test. The start time of each step is labeled on the graph.}
\label{fig:Dependence of the light output on the elapsed time of the irradiation test}
\end{figure}

\begin{figure}[H]
\centering\includegraphics[width=9cm]{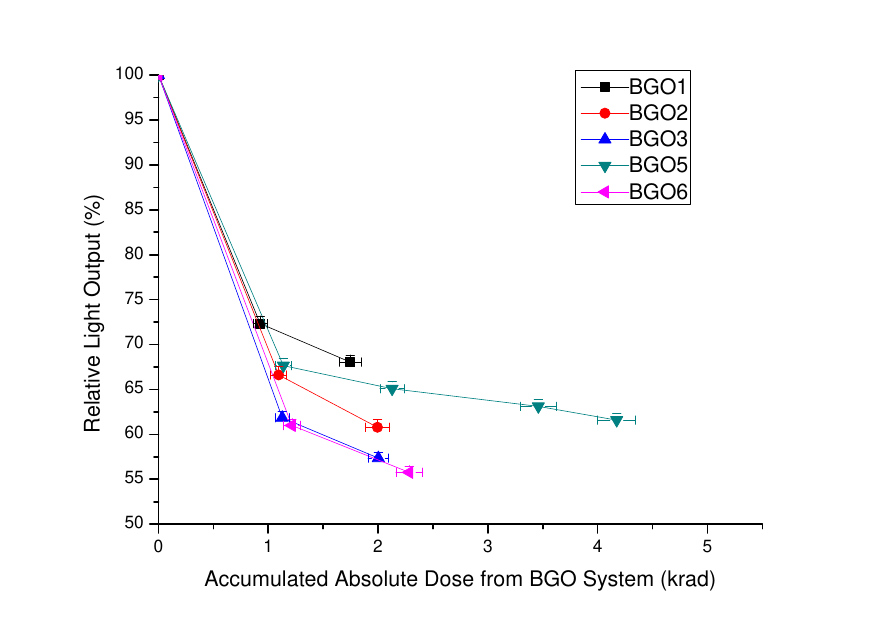}

\caption[Dependence of the relative light output on the accumulated absolute
dose from BGO system.]{Dependence of the relative light output on the accumulated absolute
dose from BGO system. The first point correspond to the value prior
to the radiation exposure and is normalized to one.}
\label{fig:Dependence of the relative light output on the accumulated absolute dose from BGO system}
\end{figure}

\begin{figure}[h]
\centering\includegraphics[width=9cm]{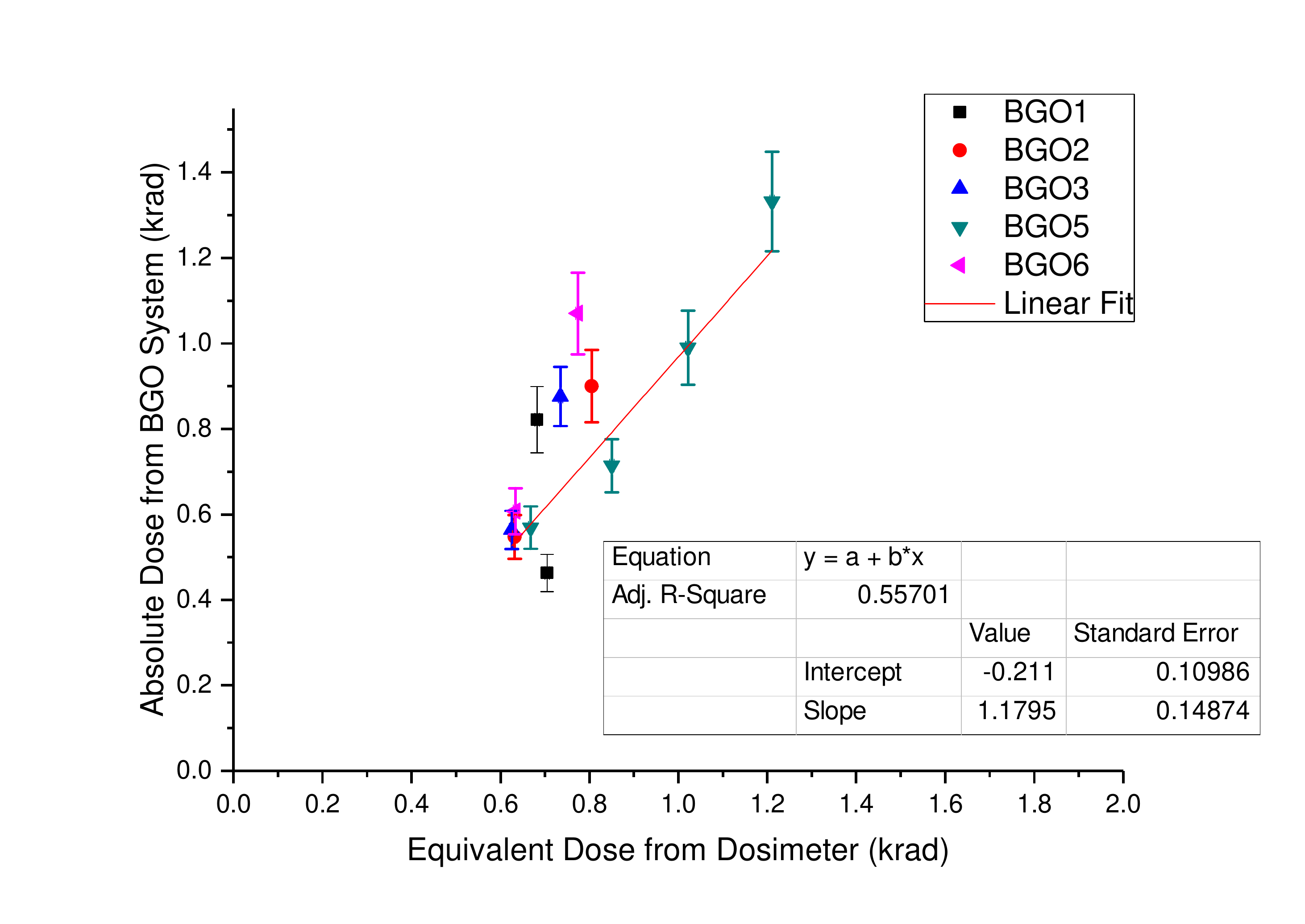}

\caption[Linearity plot: absolute dose from BGO system vs. equivalent dose
from dosimeter.]{Linearity plot: absolute dose from BGO system vs. equivalent dose
from dosimeter.}
\label{fig:Linearity plot: absolute dose from BGO system vs. equivalent dose from dosimeter}
\end{figure}

\begin{figure}[H]
\centering\includegraphics[width=9cm]{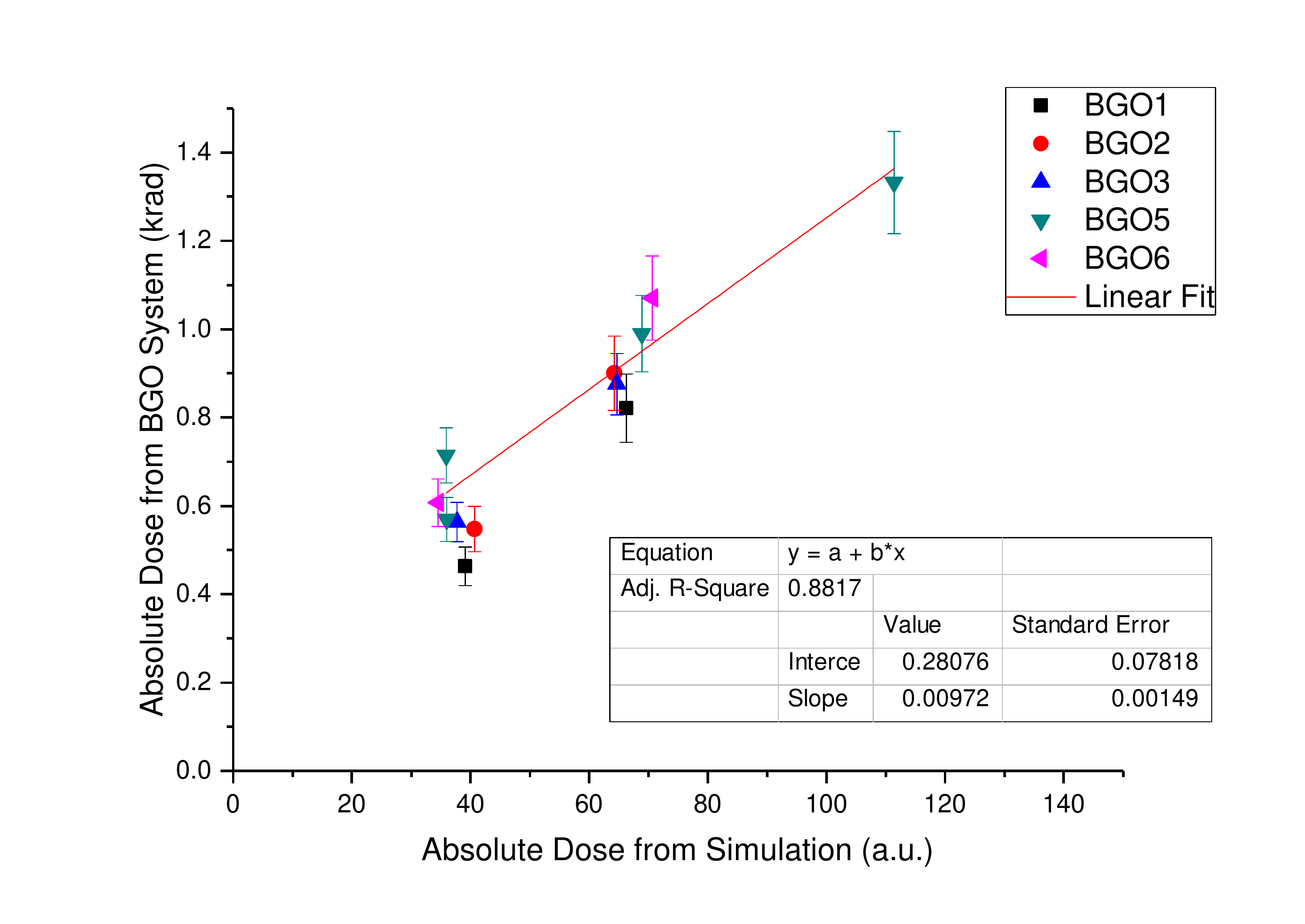}

\caption[Linearity plot: absolute dose from BGO system vs. absolute dose from
simulation.]{Linearity plot: absolute dose from BGO system vs. absolute dose from
simulation.}
\label{fig:Linearity plot: absolute dose from BGO system vs. absolute dose from simulation}
\end{figure}

The results of the irradiation are relatively monotonous decreases
of the light outputs with the elapsed time. The kinetics of the light
outputs under irradiation is shown in figure \ref{fig:Dependence of the light output on the elapsed time of the irradiation test}.
The light outputs increase as the decrease of $L$, the distance between
the source and the BGO crystals. As the crystals are irradiated transversely,
the light outputs severely saturate for BGO-1, and decrease from BGO-1
to BGO-6 gradually. The extra light outputs due to the bare crystal
in the first step are less than 3\%. The overflow data and the extra
light outputs are excluded from our following analyses. The typical
change of the relative light output, depending on the absorbed $\gamma$-radiation
dose within the interval of 1\textendash 5 krad is shown in figure
\ref{fig:Dependence of the relative light output on the accumulated absolute dose from BGO system}.
The light outputs drop abruptly at 1 krad doses, reaching values close
to the saturation. The absolute dose measured by the BGO system as
a function of the dose from the dosimeters and from the simulation
are shown in \ref{fig:Linearity plot: absolute dose from BGO system vs. equivalent dose from dosimeter}
and figure \ref{fig:Linearity plot: absolute dose from BGO system vs. absolute dose from simulation},
respectively. The linearity appears very good.

We studied the dependence of light outputs on the elapsed time and
the absorbed dose, respectively, and compared the absolute doses from
the BGO system with those obtained from the dosimeters and from the
simulation, respectively. The results reveal that the exposure of
BGO crystals to $^{60}$Co $\gamma$-ray doses of 1 krad has led to
immediate light output reductions of 25\textendash 40\%, and the light
outputs further drop by 30\textendash 45\% after the crystals receive
doses of 2\textendash 4 krad. The BGO crystals appear to receive more
doses when irradiated transversely. A salient $\gamma$-ray shielding
ability of the BGO crystals is also observed. The absolute dose from
the BGO system is consistent with the simulation, and is estimated
to be about 1.18 times the equivalent dose. 

Our findings agree with those of the survey of the scintillation properties
of the BGO crystals for EFC at Belle \cite{RN428}, which stated that
the crystals show the light yield drops about 25\textendash 50\% after
receiving 1 krad dose and remain stable afterwards. Our results also
correspond with the reported features of the N-type BGO crystals grown
by the LTG Cz technology \cite{RN426,RN427,RN424}, which revealed
that the N-type crystals show strong abrupt degradation to 35\textendash 50\%
upon irradiation and slow relaxation, and may become yellow tinted
after receiving 1 krad $\gamma$-radiation doses.

The irradiation study proves that the BGO system is able to monitor
the background dose rate in real time under extreme high radiation
conditions. The results of the irradiation study also agree with the
those of the previous studies performed on the RH BGO crystals grown
by the LTG Cz technology. This demonstrates the success of the BGO
detector system, and ensure that the data provided by this system
for BEAST II are reliable.

\section{Conclusions}

The BGO background monitor designed by the NTUHEP group was calibrated
and tested. The overall radiation sensitivity of this system is estimated
to be $(2.20\pm0.26)\times10^{-12}$ Gy/ADU with the standard 10 m
fibers for transmission and the MAPMT operating at 700 V.

The results of the irradiation study reveal that the exposure of BGO
crystals to $^{60}$Co $\gamma$-ray doses of 1 krad has led to immediate
light output reductions of 25\textendash 40\%, and the light outputs
further drop by 30\textendash 45\% after the crystals receive doses
of 2\textendash 4 krad. Our findings agree with those of the previous
studies on the RH BGO crystals grown by the LTG Cz technology \cite{RN428,RN426,RN427,RN424}.
The absolute dose from the BGO system is also consistent with the
simulation, and is estimated to be about 1.18 times the equivalent
dose. These results prove that the BGO system is able to monitor the
background dose rate in real time under extreme high radiation conditions.
This study concludes that the BGO system is reliable for the beam
background study in BEAST II.

\appendix
\acknowledgments

We are very grateful for the help from the INER people during the
irradiation test. This work is supported by the Ministry of Science
and Technology (Taiwan) under the grant MOST 103-2112-M-002-017-MY3.

\bibliographystyle{JHEP}
\nocite{*}
\bibliography{References}

\end{document}